\documentclass[letterpaper,twocolumn,10pt]{article}
\usepackage{usenix-2020-09}

\usepackage{tikz}
\usepackage{amsmath}
\usepackage{amssymb}
\usepackage{graphicx}
\usepackage{subfigure}
\usepackage{subcaption}

\usepackage{xspace}
\usepackage{paralist}
\usepackage{soul}
\usepackage{enumitem}
\usepackage{booktabs}

\usepackage{algpseudocode}
\usepackage[ruled,linesnumbered,vlined]{algorithm2e}
\usepackage{multirow}
\usepackage{makecell}
\usepackage{hyperref}
\usepackage{comment}

\hypersetup{
    colorlinks=true,
    linkcolor=blue,
    filecolor=magenta,      
    urlcolor=blue,
}

 \usepackage{etoolbox}
 \makeatletter
 \patchcmd{\maketitle}
 	{\@maketitle}
 	{\@maketitle}
 	{}
 	{}
 \makeatother

\newcommand{\fan}[1]{\textcolor{red}{(#1 --Fan)}\xspace}
\newcommand{\yuming}[1]{\textcolor{blue}{(#1 --yuming)}\xspace}

\definecolor{brickred}{rgb}{0.8, 0.25, 0.33}
\definecolor{BRICKRED}{rgb}{0.8, 0.25, 0.33}
\definecolor{BLUE}{rgb}{0, 0, 255}

\usepackage[compact]{titlesec}

\begin{document}

\newcommand{\OurSys}{\textsc{SPire}\xspace}

\newcommand{\PartitionFactor}{Partition Density\xspace}

\newcommand{\Baseline}{\textsc{Share-nothing}\xspace}

\definecolor{richgold}{RGB}{212, 175, 55}

\date{}





\title{Scalable Distributed Vector Search via Accuracy Preserving Index Construction}


\author{
{\rm Yuming Xu$^1$\thanks{Work done during his internship at Microsoft in 2023.}, Qianxi Zhang$^2$, Qi Chen$^2$, Baotong Lu$^2$, Menghao Li$^3$, Philip Adams$^3$,}\\
{\rm Mingqin Li$^4$, Zengzhong Li$^5$, Jing Liu$^2$, Cheng Li$^1$, Fan Yang$^2$}
\\ \vspace{2mm}
$^1$University of Science and Technology of China, $^2$Microsoft Research, $^3$Microsoft\\
$^4$Shopify, $^5$Microsoft AI and Research
}

\maketitle

\begin{abstract}

Scaling Approximate Nearest Neighbor Search (ANNS) to billions of vectors requires distributed indexes that balance accuracy, latency, and throughput. Yet existing index designs struggle with this tradeoff. This paper presents \OurSys{}, a scalable vector index based on two design decisions. 
First, it identifies a balanced partition granularity that avoids read-cost explosion. Second, it introduces an accuracy-preserving recursive construction that builds a multi-level index with predictable search cost and stable accuracy.
In experiments with up to 8 billion vectors across 46 nodes, \OurSys{} achieves high scalability and up to 9.64× higher throughput than state-of-the-art systems.

\end{abstract}

\section{Introduction}

Modern systems increasingly rely on high-dimensional vector indexes for Approximate Nearest Neighbor Search (ANNS), an important AI workload~\cite{gao2024arxiv, fan2024kdd, liu2024retrieval, chen2024magicpig, chatgptplugin, Taobao, Wei2020VLDB, VEARCH2018MIDDLEWARE, gao2024arxiv, shi2024acl, Borgeaud2021ICML, huang2020kdd, Suchal2010Full, zhang2018kdd}. A vector index usually connects nearby vectors to form a graph, and queries traverse the graph to locate approximate neighbors. Because ANNS trades exactness for efficiency, maintaining high query \emph{accuracy} is critical (e.g., recalling >90\% of neighbors). Consequently, a high-quality vector index typically preserves sufficiently dense graph connectivity to ensure broad coverage.

Scaling ANNS to billions of vectors requires distributing the index across server nodes and utilizing on-disk storage, which requires index sharding. As high accuracy demands dense index graph connectivity, the sharding makes cross-node links unavoidable. Unfortunately, these links introduce substantially higher latency than local traversal~\cite{adams2025distributedann, gotte2024arxiv}. 

To reduce cross-node links, a common approach is to cluster neighboring vectors into partitions, elect partition centroids, and connect nearby centroids. A query then traverses toward relevant centroids before locating approximate neighbors within their associated partitions. Several ANNS systems stack such clustering hierarchically to further control cross-node links, so as to improve scalability with low latency~\cite{ChenW21NIPS,Wei2020VLDB,PineconeServerless,su2024vexless,xu2025harmony}.

However, vector clustering introduces fidelity loss. A centroid may be a poor representative of its partition, especially for vectors near partition boundaries. To compensate for this fidelity loss, queries often have to probe additional partitions to recover accuracy, which increases vector reads (with extra CPU cycles and disk IOs) and thus reduces overall system throughput. 

This tension creates a fundamental tradeoff between query \emph{latency} (dominated by cross-node communication) and system \emph{throughput} (dominated by vector reads), given an accuracy target. Despite extensive prior work on vector indexes, this tradeoff in scalable hierarchical ANNS indexes remains underexplored and poorly understood. As a result, existing distributed ANNS systems often suffer inflated latency or constrained throughput~(\S\ref{background}).

To address these limitations, this paper presents \OurSys{}, a hierarchical vector index system designed for high scalability. \OurSys{} distinguishes itself through the following two design decisions. 

First, given a target query accuracy, \OurSys{} identifies a balanced partition granularity that avoids significant throughput penalties. We observe that, under a fixed accuracy, the average vector read cost of a query, proportional to throughput penalties, remains largely insensitive to increases in partition granularity when each partition contains only a few vectors. But the read cost grows sharply once the granularity exceeds this range. This is because the read cost is inversely proportional to the partition density, a metric we introduce to quantify the partition granularity~(\S\ref{design:density}). \OurSys{} therefore selects the granularity just before this inflection point.


Second, because this balanced granularity may still be too fine-grained and induce excessive cross-node communication, \OurSys{} employs an accuracy-preserving recursive index construction strategy that builds a multi-level hierarchical index. Cross-node traversal occurs only across levels. Each level in the hierarchy is treated as a single-level vector index, allowing \OurSys{} to decompose the end-to-end accuracy preservation into a sequence of level-by-level index construction tasks. The first level is built using a conventional approach while adopting the balanced partition granularity to satisfy the global accuracy target. The second level applies the same procedure to the centroids of level-1 partitions, but adjusts construction to preserve the \emph{end-to-end} accuracy rather than optimizing for the level-2 accuracy alone. 

The above process repeats until the top-level index fits within the memory budget of a single server. This recursive, level-by-level construction provides a fixed search cost per level, resulting in stable and highly predictable end-to-end search behavior across scales and significantly simplifying system design and implementation.


Guided by these two design decisions, we implemented \OurSys{} in 6,000 lines of C++ code. Extensive experiments show that \OurSys{} is the first system to achieve high throughput scalability for high-dimensional vector search. It outperforms state-of-the-art systems by up to 9.64× in peak throughput in production-scale deployments (up to 8 billion vectors across 46 nodes), while maintaining lower average and tail latency at all scales. The system saturates SSD (I/O-bound) while using only a fraction of network (<30\%) and CPU (<40\%) capacity, demonstrating strong headroom for further scaling.

Operationally, \OurSys{} is simple to deploy. Only the top-level index is kept in memory, while lower-level indexes as well as vectors are offloaded to SSDs. This design keeps the compute tier stateless: the in-memory index is replicated across nodes and can be reconstructed from the SSDs in case of catastrophic failures, enabling elastic scaling and simplified fault recovery. We plan to release the \OurSys{} codebase.

\section{Background}
\label{background}
\subsection{Vector Search Systems}
Vector search, which performs similarity search over high-dimensional spaces, has become a core system component powering latency-sensitive applications such as web search~\cite{huang2020kdd}, recommendation~\cite{Suchal2010Full, zhang2018kdd}, e-commerce~\cite{Taobao,Wei2020VLDB,VEARCH2018MIDDLEWARE}, and retrieval-augmented generation (RAG)~\cite{gao2024arxiv, shi2024acl, Borgeaud2021ICML, hu2025hedrarag, ray2025metis}.

Due to the curse of dimensionality, exact nearest neighbor search is often expensive~\cite{clarkson1994algorithm}. Modern systems instead rely on vector indexes for Approximate Nearest Neighbor Search (ANNS)~\cite{Fu2019VLDB, Malkov2018TPAMI, Jang2023ATC, Suhas2019NEURIPS, ChenW21NIPS, Wei2020VLDB, wang2010semi,mu2010non,song2011multiple,weiss2009spectral,xu2011complementary}. These systems trade exact search results for high efficiency: they must deliver high recalls (often above 0.9~\cite{gao2024arxiv, bigann2023, Zhang2023NSDI}) of an ANNS, respond within millisecond-level SLOs (e.g., 10 \textasciitilde\ 20 ms)~\cite{Taobao, xu2023sosp, Zhang2023NSDI}, and sustain high query loads that routinely reach thousands of QPS~\cite{Wei2020VLDB, MILVUS2021SIGMOD, adams2025distributedann}.

Among various types of ANNS indexes, graph-based indices, such as HNSW~\cite{Malkov2018TPAMI}, dominate single-node deployments due to their efficient search performance and high accuracy.  These indices connect nearby vectors to form a densely connected proximity graph where queries use best-first search to locate top-$K$ neighbors approximately.



\subsection{Scaling Vector Search Systems}
\label{sec:distributed-search}


\noindent\textbf{Inflated latency due to index sharding.} When scaling to billions of vectors~\cite{fan2024kdd, Gan2023BinaryER}, index sharding becomes necessary. A na\"ive approach is to shard the popular graph-based vector index into multiple partitions. However, graph-based indexes like HNSW rely on dense connectivity to ensure query accuracy. Sharding a densely connected graph unfortunately introduces enormous cross-partition links which connect shards across nodes. A query traversing such a sharded global graph must issue frequent Remote Procedure Calls (RPCs), thus inflating latency.

Further complicating the situation, the query process is inherently sequential and data-dependent. A query decision for the next best traversal path is determined on the fly: the next step only becomes obvious after evaluating the distances of all currently known neighbors. The sequential nature of vector query makes conventional speedup techniques like parallelization or prefetching hardly applicable.


\begin{table}[!t]
    \centering
    \resizebox{\linewidth}{!}{
    \begin{tabular}{|c|c|c|c|c|}
    \hline
     Dataset & \multicolumn{2}{c|}{SPACEV100M} & \multicolumn{2}{c|}{SIFT100M} \\
    \hline
    Recall  & 0.9 & 0.95 & 0.9 & 0.95 \\
    \hline
    Avg Total Steps & 94.25 & 222.51 & 87.31 & 146.23 \\
    \hline
    Avg Cross-node Steps & 80.90 & 209.15 & 73.43 & 132.35 \\
    \hline
    P99 Cross-node Steps & 140 & 310 & 92 & 146 \\
    \hline
    \end{tabular}
    }
    \caption{Cross-node communication dominates HNSW traversal cost. Average total and cross-node traversal steps show that when an HNSW index is sharded across five nodes, remote traversals consistently constitute the majority ($\mathrm{>80\%}$) of all search operations, causing high query latency.}
    \label{tab:background:crossnode}
\end{table}

\begin{figure}[!t]
    \centering    
        \includegraphics[width=1\linewidth]{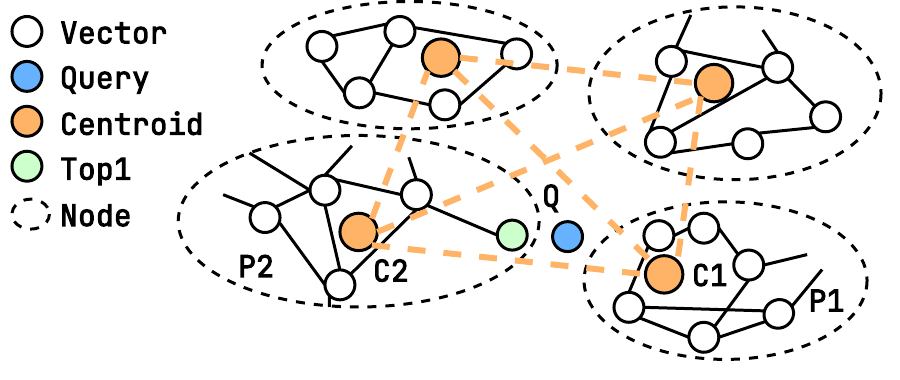}
    \caption{Partitioned routing reduces communication but necessitates excessive partition probing to mitigate Fidelity Loss, severely limiting throughput. (Query $Q$ is directed to partition $P_1$ because it is closer to centroid $C_1$. However, the true nearest neighbor resides in partition $P_2$ (closest to $C_2$)).}
    \label{fig:background:representativenessLoss}
\end{figure}

This challenge is severe and quantifiable. Table~\ref{tab:background:crossnode}
shows that in a 100M-vector HNSW index sharded across five nodes, cross-node traversals constitute over 80\% of the total search steps. This overwhelming cross-node access is orders of magnitude slower than local vector reads, translating to high latency. As a result, the p99 latency increases by two orders of magnitude to tens of milliseconds, even after adopting similar optimizations discussed in~\cite{adams2025distributedann}.
This reveals a fundamental structural limitation, where traversing in a sharded dense graph incurs significant remote communication costs, thus it is a poor fit for low-latency SLOs.




\noindent\textbf{Reduced throughput due to fidelity loss.} To reduce cross-node links, distributed vector search systems increasingly adopt a partition-based hierarchical design~\cite{Nara2014VLDB, FAISS2020, MILVUS2021SIGMOD, PineconeServerless}. 
This design employs a practical divide-and-conquer strategy: the dataset is partitioned into manageable shards, and the centroids of partitions route queries to the most relevant partitions.
Figure~\ref{fig:background:representativenessLoss} illustrates a two-level index hierarchy, adopted by systems such as DSPANN~\cite{ChenW21NIPS} and ADBV~\cite{Wei2020VLDB}. To further reduce remote communications, more recent systems like Pinecone~\cite{PineconeServerless} extend to a multi-level hierarchy, similar to SPTAG~\cite{Chen2018Github} or tree-based structures like KD-trees~\cite{KDTREE1975ACM} and R-trees~\cite{RTREE1984SIGMOD}.


However, vector clustering introduces \emph{fidelity loss}. Vectors near partition boundaries are poorly represented by their centroid.
As illustrated in Figure~\ref{fig:background:representativenessLoss}, this fidelity loss misleads query $Q$ into the wrong partition (represented by $C_1$), even though its true nearest neighbor resides in the other partition $C_2$, whose centroid is further away.
To compensate for the fidelity loss, queries have to probe more partitions to recover accuracy.
 For instance, in a large-scale DSPANN deployment, recalling >90\% of the top-5 neighbors ($\text{Recall}@5=0.9$) requires queries to search $4$ of $5$ partitions for a $1\text{B}$ index and $9$ of $46$ partitions for an $8\text{B}$ index. The extra probing incurs excessive vector data reads, consuming additional CPU cycles and IOs across multiple nodes, and this amplification of read and compute operations reduces the maximum sustainable throughput (QPS).


\subsection{Latency–Throughput Trade-off}
\label{subsec:trade}

The above two challenges of scaling vector indexes reveal a fundamental trade-off between query \emph{latency} and \emph{throughput}, given a target query accuracy. Na\"ive indexing sharding preserves high index fidelity but inflates latency; while vector clustering reduces latency but degrades fidelity, ultimately hurting throughput.


Such a tradeoff in scalable distributed ANNS indexes remains
underexplored and poorly understood. Despite extensive prior studies on vector indexes discussed above, they have suffered either high latency or limited throughput~(\S\ref{evaluation}). 

\section{\OurSys Design}
\label{design:overview}

\begin{figure}[!t]
    \centering
\includegraphics[width=1\linewidth]{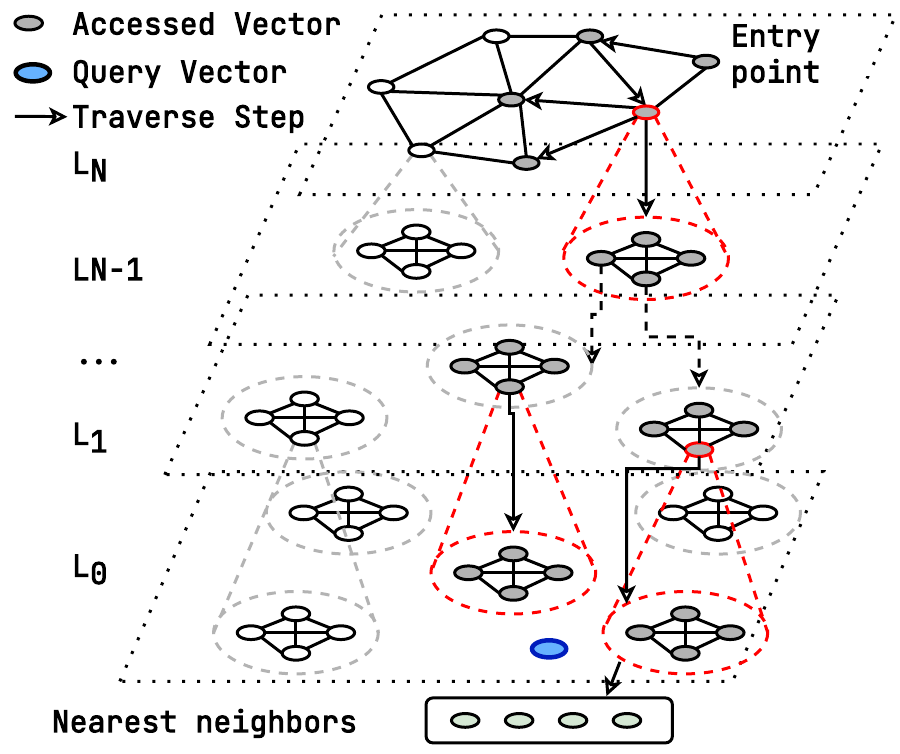}
    \caption{The index structure of \OurSys{}.
    }
    \label{fig:IndexSystemOverview}
\end{figure}

\begin{figure} [!t]
\includegraphics[width=1\linewidth]{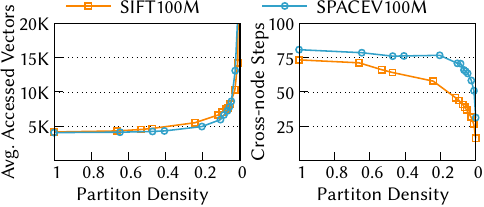}
\caption{A balanced partition granularity.
As partition density decreases (granularity coarsens), more vector reads are required to preserve the accuracy of Recall@5 (left) while cross-node steps decrease (right). An inflection point exists in the Figure (left), i.e, a balanced granularity, where accuracy is preserved without excessive vector reads and cross-node hops are reduced.}
\label{fig:Motivation}
\end{figure}

\noindent


\subsection{Overview}

We present \OurSys{} to address the above challenges and the trade-off between latency and throughput. As shown in Figure~\ref{fig:IndexSystemOverview}, \OurSys{} organizes the index into multiple levels. The root (top) level maintains an in-memory proximity graph, while the remaining levels cluster vectors into partitions stored in SSDs. The index is constructed bottom-up: we first cluster all vectors at a chosen granularity, and then recursively cluster the resulting centroids until the root level fits within the memory budget of a single machine. Consequently, each vector at level $L_x$ serves as the centroid of a partition at the adjacent lower level $L_{x-1}$, forming a 
hierarchy that enables efficient traversal. 
The partition granularity is carefully selected to avoid read amplification and enable high-throughput scalability (\S\ref{design:density}).

A query traverses the hierarchy from the root level $L_N$ down to the leaf level $L_0$, progressively narrowing the search space before returning the final results (\S\ref{design:hierarchy}). At each level, the query identifies the top-$m$ nearest vectors and then descends to the next level by fetching the corresponding $m$ partitions, within which it again computes the top-$m$ candidates for further traversal. The search process resembles that of a traditional B\texttt{+}-tree, with the key distinction that \OurSys{} explores multiple ($m$) partitions at each step rather than a single child node to preserve the accuracy in the high-dimensional space.

The query latency of \OurSys{} is inherently bounded by the height of the hierarchy, as the $m$ partitions at one level can be processed in parallel in one network round trip. 
Our evaluation (\S\ref{sec:eval_simulator}) shows that indexing 1024 billion vectors requires only 6 levels with 4~GB single node memory, and 4 levels with 512~GB, yielding an end-to-end latency smaller than 20~ms.
The index design also simplifies distributed deployment: partitions are stored as objects in a distributed storage system, while only the root index is maintained on compute servers, which we detail in \S\ref{sec:impl}.

\subsection{Balanced Partition Granularity}
\label{design:density}
We observe that, under a target query accuracy, there exists a strongly non-linear relationship 
between partition granularity and the overall overhead of distributed vector search.
This behavior makes the choice of granularity a first-order design decision: 
Overly coarse partitions suppress throughput, while overly fine partitions inflate end-to-end latency. 

This is because system throughput is dominated by the number of vectors accessed during traversal, since vector fetches and distance computations consume most resources (e.g., I/Os and CPU cycles). Coarser partitions increase vector reads due to fidelity loss, thus hurting throughput.  
Query latency, on the other hand, is governed by data-dependent network round trips incurred by cross-node hops. Finer-gained partitions incur more cross-node hops, thus inflating latency.

To analyze the tradeoff, we quantify partition granularity using partition density, defined as 
\(D = \frac{\text{Number\_of\_partitions}}{\text{Number\_of\_vectors}}\), where a lower density corresponds to coarser partitions. According to the previous discussion, the number of vectors accessed $c$ is inversely proportional to $D$, i.e., $c \propto \frac{1}{D}$~\footnote{We assume the number of vectors in a partition $\ll \text{Number\_of\_vectors}.$}.

To illustrate this relation, we vary $D$ and measure the cost of vector reads and cross-node hops required to sustain a target accuracy of $recall@5 = 0.9$.
For each density $D$, we cluster all vectors via $k$-means, construct a proximity-graph index over the cluster centroids, and distribute the graph across five servers using spatial locality to reduce cross-node connections~\cite{CoTra}.
A partition density of one degenerates into a pure proximity-graph index, as each partition contains exactly one vector. 
The results in Figure~\ref{fig:Motivation} cover the design space used by prior distributed vector indexes~\cite{ChenW21NIPS,CoTra,adams2025distributedann}, enabling a systematic view of the underlying tradeoff.

\noindent\textbf{Impact on accessed vectors.} 
Figure~\ref{fig:Motivation} (left) shows that as density decreases and thus partitions coarsen, the number of accessed vectors remains low and stable initially. 
This is because centroids represent their partitions with high fidelity, allowing queries to precisely identify target partitions without read amplification. 
Due to that read cost is inversely proportional to partition density, the cost increases sharply when density drops below a threshold (around 0.1), at which point the reduced fidelity forces the query to traverse more partitions, where the number of vectors within a partition also grows. This indicates that partitioning does not necessarily hurt throughput, so long as partition density stays above this fidelity threshold.

\noindent\textbf{Impact on cross-node traversal.} 
In contrast, Figure~\ref{fig:Motivation} (right) shows that cross-node hops in the proximity graph decrease gradually as the partition density decreases. Nevertheless, due to the dense connectivity typical of proximity graphs, cross-node hops largely persist even at very low densities. For example, on SPACEV with a density of $0.001$, the number of cross-node hops reduces by only 52\% compared to a pure graph index. This shows that single-level index partitioning may not be an effective way to reduce cross-node hops.

\OurSys{} decides to select a granularity that preserves throughput -- critical for scalable serving -- while reducing cross-node hops. 
Intuitively, it is the coarsest granularity at which the vector-access cost remains comparable to that of a high-quality graph index. We therefore select the density at the inflection point in Figure~\ref{fig:Motivation} (left), which is just before fidelity degradation forces excessive partition probes.
We refer to this operating point as balanced granularity.

The evaluation in \S\ref{sec:eval:balanced_granularity} shows that this balanced point consistently emerges across datasets, though its exact value requires profiling. Empirically, a density around 0.1 works robustly across the datasets we tested. Since cross-node graph traversal for the centroids may still occur at this granularity (66.87 and 45.88 on SIFT100m and SPACEV100m in Figure~\ref{fig:Motivation}), we address it using a hierarchical indexing strategy described in \S\ref{design:hierarchy}.

\subsection{Accuracy-Preserving Hierarchy}
\label{design:hierarchy}

The indexing and search over centroids can be viewed as another \textit{ANNS problem}, albeit at a smaller scale than the base vectors. This observation motivates \OurSys{} to recursively apply the balanced granularity to the generated centroids until the root layer fits within the memory budget of a single machine. As a result, the number of data-dependent network round trips per query is bounded by the number of indexing levels, 
constraining the end-to-end serving latency.

Algorithm~\ref{design:IndexBuilder} summarizes the index construction procedure. It first checks whether the dataset fits within the memory budget of a single server. If so, \OurSys{} builds an in-memory proximity graph index and returns it (Lines 2--4). Otherwise, \OurSys{} clusters the vectors into partitions using the discovered balanced granularity (Lines 5--6). 
Partitions are stored in SSDs to save memory cost while 
the resulting centroids are passed recursively into the build function to determine whether they can be organized by a single-server graph index or require further clustering (Line 7). This process yields a hierarchical index in which each level maps centroids to the partitions in the level below, ultimately associating the root-level centroids with the base vectors at the leaves. 

By applying the recursive index building, the hierarchy height is \( \log S \), where $S$ is the number of base vectors. 
Our evaluation demonstrates that the hierarchy is typically shallow, with four levels for 8B to 32B vectors and at most six levels for 1024B vectors with 4~GB memory budget, and the numbers reduce to three for 8B to 32B and for for 1024B for 512~GB memory budget.

Since partitions in each level are constructed recursively, we must tune search parameters in an end-to-end manner to preserve accuracy. 
Each non-leaf level determines the number of partitions to fetch from its child level; thus, an index with $N+1$ levels requires tuning 
$N$ parameters. To simplify the tuning process and ease the deployment in practice, \OurSys{} enforces that all levels fetch the same number of partitions for a query. 
Since each upper level operates on fewer vectors than the level below, using identical search budgets guarantees that upper levels achieve higher accuracy. 
This property is crucial for maintaining end-to-end accuracy; 
if an upper level is inaccurate, the correct results cannot be recovered downstream.

Thanks to the balanced granularity, centroids faithfully represent the underlying partitions. Thus, adding hierarchy levels does not induce excessive additional reads to maintain the same end-to-end accuracy. 

\noindent\textbf{Search operation.}
As illustrated in Figure~\ref{fig:IndexSystemOverview}, the query first identifies the top-$m$ nearest centroids at the root level. 
These centroids correspond to the $m$ partitions in the next level, which are fetched in parallel across servers to minimize latency.
Given that each partition contains tens of vectors at a partition density of 0.1, \OurSys{} performs a brute-force search within each partition 
to identify another top-$m$ nearest vectors to the query. 
The top-$m$ results are then propagated to the next level, and the process repeats until reaching the leaf level $L_0$
where the final top-$k$ vectors are produced.
The search cost is bounded by \( O(m\log N) \): the hierarchy height is \( \log N \) and each level requires search $m$ partitions.

\noindent\textbf{Index updates.}
\OurSys{} naturally supports insertion and deletion by incorporating existing ANNS update techniques~\cite{xu2023sosp,FreshDiskANN2021,fast26odinann}. 
Updates first occur at the leaf partitions and propagate upward as needed. \OurSys{} adopts the split-and-merge operations 
from LIRE~\cite{xu2023sosp} to maintain partition quality, while updates to the root-level proximity graph follow 
established graph-update strategies~\cite{FreshDiskANN2021,fast26odinann}.


\begin{algorithm}[!t]
\small
\caption{\OurSys Build Procedure}
\label{design:IndexBuilder}
\BlankLine
\SetKwFunction{BuildLevels}{BuildLevels}
\SetKwFunction{DetermineDensity}{DetermineDensity}
\SetKwFunction{MeetCapacityBudget}{MeetCapacityBudget}
\SetKwFunction{BuildRootIndex}{BuildRootIndex}
\SetKwFunction{BuildInnerPartitionIndex}{BuildInnerPartitionIndex}
\SetKwFunction{Partitioning}{Partitioning}
\SetKwFunction{Length}{Length}
\SetKwProg{Fn}{Function}{:}{}
\Fn{\BuildLevels{$CapacityBudget$ B, $Data$ D}}{
    \If{\MeetCapacityBudget{$B,D$}} {
        $Graph \gets$ \BuildRootIndex{$D$}\;
        \Return $Graph$\;
    }
    $Density\ R \gets$ \DetermineDensity{$D$}\;
    
    \tcp{Each $C$ maps to a $P$, $Ps[C] = P_c$}
    $Partitions\ Ps$, $Centroids\ Cs \gets$ \Partitioning{$D,R$}\;
    \tcp{Recursive indexing for upper levels}
    $UpperLevels\ Ls \gets$ \BuildLevels{B, Cs}\;
    $SPire\ S \gets \{Ls, Ps\}$\;
    \Return $S$\;
}

\end{algorithm}

\section{System Implementation}
\label{sec:impl}
\noindent
\OurSys{} comprises four key components: a parallel index construction module, a disaggregated index store, a stateless query execution engine, and a near-data processing module to minimize cross-tier data movement. The implementation enables separate scaling of storage and compute, effective load balancing, 
and robust operation. The system is implemented in about 6,000 lines of additional C++ code based on SPTAG, a popular open-source vector store~\cite{Chen2018Github}.

\subsection{Parallel Index Construction}
\noindent
Bottom-up index construction is costly because the bottom level contains a large number of clusters (proportion of the total vector number), leading to a clustering complexity of O($S^2$). \OurSys implements a five-stage process to build the index in parallel across $M$ worker nodes, reducing the complexity to O($(S/M)^2$). In practice, $S/M$ is typically constant. 

\noindent\textbf{Stage 1: Sampling-based granularity selection.}
To determine the balanced partition granularity ($G$), $\OurSys$ employs a sampling-driven iterative search mechanism.
A randomly sampled subset of the vector set is built to represent the overall vector distribution. For a large-scale vector set, we typically sample around one million vectors, drastically reducing the computational and time overhead. 
Empirically, the effective partition density is greater than $0.001$, corresponding to a range of 1 to 1,000 vectors per partition.
The search process begins by evaluating the finest granularity, where each partition contains a single vector, to establish the baseline query cost of a local graph index. Subsequently, we iteratively explore coarser granularities using a binary search over the feasible range. In each iteration, we cluster sampled vectors with the granularity $G$ and measure the number of accessed vectors required to reach a target recall (e.g., 0.9). The process halts when a sharp increase in accessed vectors, indicating the inflection point as detailed in~\S\ref{design:density}.

\noindent\textbf{Stage 2: Partitioning with boundary vector replication.}
The full vector set is first partitioned across the $M$ nodes using the distributed $k$-means algorithm. To 
reduce recall loss at partition boundaries, vectors near the $k$-means boundaries are replicated to neighboring nodes, ensuring the subsequent local clustering captures necessary cross-boundary relationships.

\noindent \textbf{Stage 3: Parallel local clustering.}
After partitioning the vector set and replicating boundary vectors, each node $i \in [1, M]$ processes its local partition $\mathcal{P}_i$ independently. All vectors within $\mathcal{P}_i$ are clustered using the discovered balanced granularity $G$, producing Level-0 centroids. As clustering relies only on local data, this phase runs in parallel without cross-node synchronization.

\noindent \textbf{Stage 4: Global partition shuffling.}
To combine all local partitions into global partitions and mitigate hot-spots under skewed workloads, we assign each partition a global identifier based on its centroid global vector ID and uniformly shuffle the partitions. 
During shuffling, partitions with the same global identifier are merged for boundary alignment.

\noindent \textbf{Stage 5: Recursive hierarchy construction.}
After building the current level index, a designated coordinator collects necessary metadata of the centroids in each node and initiates Stages 1--4 for the next level index, and the process continues until a single node can accommodate all top-level centroids. At the top level, a graph-based index is constructed over the final centroids. This iterative design transforms data-intensive clustering into a parallel process to accelerate index construction, while ensuring each hierarchical level is well-tuned.


\subsection{Disaggregated Efficient Index Store}
\label{impl:index_store}
The index store is a distributed storage layer separate from the compute tier. It stores the hierarchical index in SSDs and exposes a lightweight RPC interface to compute nodes. Each partition is kept in the index store as an independent object containing a sequence of vector entries along with their vector IDs and associated contents. Partitions are assigned unique Partition Identifiers ($\mathit{P_{ID}}$), implicitly defined as the vector IDs of their parent centroids. 

The index store relies on a hash function $H$ (e.g., consistent hashing) over $\mathit{P_{ID}}$ for partition placement. The hash function uniformly distributes partitions across the $M$ storage nodes, mitigating hot-spots under skewed workloads and facilitating partition routing through $H(\mathit{P_{ID}}) = storage\_node\_ID$. 



\noindent \textbf{Near-data processing.}
To further reduce the large data transmission overheads, 
instead of shipping partitions to the compute nodes, storage nodes execute partition navigation, compute vector--query distances in parallel, and perform candidate aggregation to return only a compact set of top-$N$ results. These operations are exposed to compute nodes via RPCs like \texttt{GetPartitionResult}. The returned results use a compact representation, e.g., IDs occupy 8\,B and distances consume 4\,B, keeping the response under 6\,KB for $N < 512$, compared with hundreds of KB of raw vectors in baseline systems.


\subsection{Stateless Query Execution Engine}

The query execution engines run on compute nodes to execute queries. A single engine instance manages the entire process of each query, calling the index store to traverse every hierarchy level. The engine caches the top-level index to reduce latency.

The engine is stateless, and critical data are persisted to the index store. Therefore, the compute nodes running the engine can be elastically scaled up or down easily, allowing \OurSys{} to adapt quickly to workload fluctuations. 


To improve performance, the query engine batches multiple requests to storage nodes into asynchronous \texttt{GetPartitionResult} RPCs, who performs near-data computation (\S\ref{impl:index_store}).

From top to bottom, the engine executes a query level-by-level. At each level, it fetches the relevant partitions from the index store, merges and ranks the returned candidates, and uses the vector IDs of the candidates to determine the $\mathit{P_{ID}}$s of the next-level partitions to proceed further. This process repeats until the bottom level is reached with the finalized top-$k$.


\subsection{Deployment and Robust Operation}
\noindent $\OurSys$ is engineered for elasticity, cost-effectiveness, and robustness. These operational benefits stem from the \OurSys{}'s stateless query engine, disaggregated architecture, and deliberate simple placement strategy.

\noindent\textbf{Cost-effective and elastic scalability.}
Decoupling the query execution engine (compute) from the index store (storage) enables scaling compute and storage tiers independently without over-provisioning each other, thus achieving cost effectiveness.  
The query engine's stateless design allows compute nodes to be elastically scaled up or down easily. This ensures rapid adaptation to workload fluctuations and improves resource utilization.


\noindent\textbf{Fault tolerance.}
The stateless engine makes fault tolerance easy: queries can be transparently routed to any healthy node without worrying about state consistency. The index store ensures durability through replication and relies on randomized partition placement to alleviate hotspots and maintain load balance, even under failures.
Together, these design choices allow \OurSys{} to preserve high availability and stable performance in the presence of compute or storage node outages. 

\section{Evaluation}
\label{evaluation}
\noindent This section evaluates \OurSys{} in scalability, hierarchical behavior, and system efficiency:

\begin{itemize}[leftmargin=*, nosep]
    \item \textbf{Performance and scalability (\S\ref{sec:eval_goodscalability}).} 
    \OurSys{} achieves high scalability up to 8B vectors and delivers up to $9.64\times$ higher throughput than baselines, while consistently maintaining lower tail latency. 
    \item \textbf{Simulation at extreme scale.(\S\ref{sec:eval_simulator}).} 
    Our analytical model and simulator show that at extreme scale (1024B), \OurSys{} achieves high throughput scalability with disk I/Os as the dominant bottleneck.
    \item \textbf{Existence of balanced granularity (\S\ref{sec:eval:balanced_granularity}).} 
    We empirically validate the existence of a balanced granularity that remains consistent across datasets with diverse modalities, distance metrics, and data scales.

    \item \textbf{Effect of \OurSys{} design choices (\S\ref{sec:eval_hierarchy_benefits}).} 
    \OurSys{}'s balanced granularity and accuracy-preserving hierarchy significantly reduce computational overhead compared to alternative approaches.

    \item \textbf{Micro analysis (\S\ref{sec:eval_ablationstudy}).} 
    The hierarchical index introduces minimal storage overhead. Near-data execution and hash-based placement effectively mitigate network latency, compute imbalance, and hot-spot contention.
\end{itemize}

\begin{figure*}[!t]
    \centering
    \subfigure[Peak throughput.]{
        \includegraphics[scale=1]{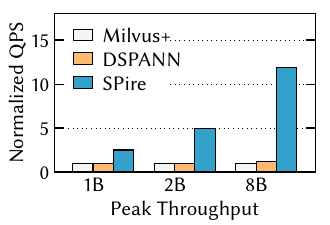}
        \label{fig:evaluation:peakThroughput}
    }
    \subfigure[Latency at peak throughput (all satisfy latency requirement).]{
        \includegraphics[scale=1]{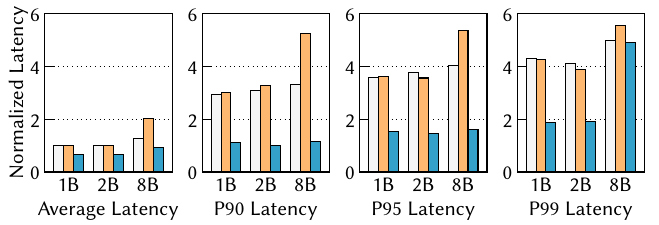}
        \label{fig:evaluation:peakLatency}
    }
    \caption{End-to-end performance across data sizes in production workload. \OurSys{} achieves high scalability, maintaining higher throughput and lower latency than all baselines.}
    \label{fig:evaluation:scalability}
\end{figure*}

\subsection{Experimental Setup}
\noindent\textbf{Platform.} We conduct experiments on two environments:
(1) Production cluster: a 46-node deployment used for the scalability experiments in \S\ref{sec:eval_goodscalability}.
(2) Standard cloud VMs: Azure Lsv3 instances~\cite{Azurelsv3} (16 vCPUs, 128~GB RAM, 2$\times$1.92~TB NVMe SSDs), used for all other evaluations.
We co-locate compute and storage components on the same physical nodes with statically partitioned resources.

\noindent\textbf{Datasets.} We evaluate \OurSys{} using the benchmark suite in Table~\ref{tab:datasets}, spanning multiple modalities (Image, Text, Web), similarity distance ($L_2$, Cosine, IP), and dataset scales (10M to 8B). Notably, we include \textit{Production-8B}, a real-world web search dataset, to demonstrate performance at billion-scale under realistic distributions.

\noindent\textbf{Baselines.} We compare \OurSys{} against three representative partitioning strategies. 
\begin{itemize}[leftmargin=*, nosep]
    \item \textbf{Milvus+:} 
    Represents the na\"ive partitioning approach used in systems like Milvus~\cite{MILVUS2021SIGMOD} and NSG~\cite{Fu2019VLDB}. Data is partitioned randomly across nodes, necessitating a scatter-gather search across all partitions. We re-implemented this strategy to ensure scalability on our 8B-vector dataset.
    \item \textbf{DSPANN:} A clustering-based partitioning strategy used in production systems. 
    ADBV~\cite{Wei2020VLDB} follows the same distribution strategy. 
   Data is partitioned via k-means and merged to balance sizes~\cite{ChenW21NIPS}.
    Following the configuration in~\cite{adams2025distributedann}, each partition is capped at 200M vectors, resulting in 46 partitions for the 8B dataset where one server contains one partition. 
    For a fair comparison, all systems employ 46 servers on the 8B dataset. 
    \item \textbf{Pinecone*:} A balanced hierarchical clustering approach introduced by  Pinecone~\cite{PineconeServerless} that subdivides large partitions in a top-down way to enforce uniform leaf sizes. Since Pinecone is closed-source, we re-implemented its design, storing leaf partitions on disk and keeping internal structures in memory to balance memory usage.
\end{itemize}

\noindent\textbf{Metrics.} Throughout the experiments, we report:
(1) Recall@$k$ (standard accuracy metric~\cite{bigann2023});
(2) Peak Throughput (QPS under accuracy and hardware constraints); and (3) Latency (average and P99).

\begin{table}[!t]
\centering
\small
\setlength{\tabcolsep}{3pt} 
\resizebox{\linewidth}{!}{
\begin{tabular}{lcccccc}
\toprule
\textbf{Dataset} & \textbf{Dim} & \textbf{Size} & \textbf{\makecell{Query\\Size}} & \textbf{Distance} & \textbf{Type} & \textbf{Source} \\ 
\midrule
SIFT1B~\cite{Sift}      & 128  & 1B   & 10k    & L2     & UInt8 & Image    \\
SPACEV1B~\cite{Spacev}    & 100  & 1B   & 29k    & L2     & Int8  & Web Doc  \\
DEEP1B~\cite{Deep}      & 96   & 1B   & 10k    & L2     & FP32  & Image    \\
OpenAI~\cite{OpenaiDataset}      & 1536 & 5M   & 1k     & Cosine & FP32  & Text     \\
Cohere~\cite{CohereDataset}      & 768  & 10M  & 1k     & Cosine & FP32  & Text     \\
BIOASQ~\cite{krithara2023bioasq}      & 1024 & 10M  & 3k     & Cosine & FP32  & Text     \\
Laion~\cite{schuhmann2022laion}       & 768  & 100M & 1k     & L2     & FP32  & Img/Text  \\
Text~\cite{TextDataset}        & 200  & 10M  & 100k   & IP     & FP32  & Text     \\
Production  & 384  & 8B   & 16k    & L2     & UInt8  & Web Doc  \\ 
\bottomrule
\end{tabular}
}
\caption{Overview of datasets.}
\label{tab:datasets}
\end{table}

\subsection{Performance and Scalability}
\label{sec:eval_goodscalability}

\noindent We evaluate \OurSys{} on the Production8B dataset at three scales: 1B, 2B, and 8B, comparing throughput and latency with the baselines. 
All systems were tuned to achieve Recall@5 = 0.9 to ensure fair comparison.
Pinecone* results are excluded because it did not achieve the target recall within acceptable latency on production-scale clusters. A more detailed discussion of its behavior is presented in~\S\ref{sec:eval_hierarchy_benefits}.
The results show that \OurSys{} significantly improves peak throughput while maintaining comparable latency.

We used 5 nodes for 1B vectors, 10 nodes for 2B, and 46 nodes for 8B. The partition density of \OurSys{} was set to 0.1 across all levels. \OurSys{} builds a three-level hierarchy for the 1B and 2B datasets and a four-level hierarchy for the 8B dataset. The search parameter was configured as \( N = 256 \),  i.e., each query accesses up to 256 partitions per level in bulk.


\begin{figure*}[!t]
    \centering
    \begin{minipage}[t]{0.24\linewidth}
        \centering
        \includegraphics[scale=1]{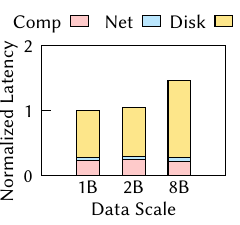}
        \caption{\OurSys{} latency breakdown. Disk accesses dominate the query overhead.}
        \label{fig:evaluation:latencyBreakdown}
    \end{minipage}
    \hfill
    \begin{minipage}[t]{0.72\linewidth}
    \centering
\includegraphics[scale=1]{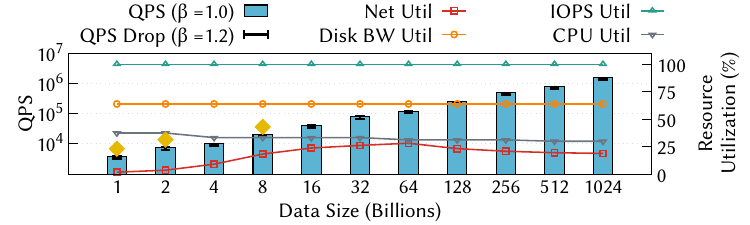}
    \caption{Simulated throughput across different data scales. $\beta$ represents the node access imbalance factor. Consistent with Figure~\ref{fig:evaluation:peakThroughput} ($\textcolor{richgold}{\blacklozenge}$ marks the validated results), disk IOPS remains the primary system bottleneck.}
    \label{fig:evaluation:simulator}
    \end{minipage}
\end{figure*}

\noindent\textbf{Throughput.} Figure~\ref{fig:evaluation:peakThroughput} compares the normalized peak throughput of \OurSys{} and baseline systems across different dataset sizes (1B, 2B, 8B). 
\OurSys{} outperforms Milvus+ by 2.50$\times$, 5.03$\times$, and 11.96$\times$, and DSPANN by 2.50$\times$, 4.91$\times$, and 9.64$\times$, respectively. 
Milvus+ requires all nodes to be queried per request, resulting in consistent but low throughput and poor scalability.
DSPANN accesses 80\%, 50\%, and 20\% of all partitions per query at 1B, 2B, and 8B due to fidelity loss, limiting throughput and scalability. 
Moreover, its coarse-grained partitioning leads to severe hot-spot contention, with the hottest node involved in 100\%, 98\%, and 80\% of all queries at 1B, 2B, and 8B. 
\OurSys{} achieves the highest throughput, enabled by a multi-level fine-grained hierarchy that reduces redundant computation and mitigates hot-spot contention.
As the dataset grows from 1B (5 nodes) to 2B (10 nodes), \OurSys{} achieves a 2$\times$ increase in peak throughput. At 8B (46 nodes), throughput improves by 4.75$\times$ relative to the 1B setting. The multi-level index preserves per-query vector accesses at each level, enabling linear throughput gains as the number of nodes doubles (from 5 to 10). When the dataset increases by ten times, an additional index level is added; for the 8B dataset, this results in an additional index level, yielding one more level than in the 1B and 2B. This extra index level introduces additional vector partition accesses and computation, resulting in slightly sublinear scaling while still significantly outperforming all baselines.

\noindent\textbf{Latency at peak throughput.} Figure~\ref{fig:evaluation:peakLatency} shows both average latency and tail latency (90th to 99th percentile) at peak throughput.
All baseline systems are designed to optimize latency and meet the application’s requirements, typically at the cost of accessing larger portions of data and incurring higher computation overhead. 
\OurSys{} consistently achieves comparable or lower latencies.
In terms of average latency, \OurSys{} reduces it by $1.53\times$, $1.51\times$, and $1.37\times$ compared to Milvus+, and $1.54\times$, $1.54\times$, and $2.21\times$ compared to DSPANN at 1B, 2B, and 8B scales, respectively.
Tail latency shows similar trends, with \OurSys{} consistently outperforming the baselines.
These results stem from \OurSys{}'s hierarchical design, where each index level processes only a subset of vectors using bulk network and disk access. 
Each level only processes a smaller set of vectors with bulk network and disk access.
As a result, the 1B and 2B datasets have nearly identical average latencies, while the 8B dataset sees a moderate increase, up to $1.37\times$, due to an additional index level. 
For p99 latency at the 8B scale, the increase mainly reflects variance in SSD I/O. Since the hierarchy remains shallow, with only six levels even at the 1024B scale, this variance is inherently bounded. The system therefore, continues to meet the application's latency requirements.

\noindent\textbf{\OurSys{} latency breakdown.} Figure~\ref{fig:evaluation:latencyBreakdown} shows the latency breakdown of \OurSys{} across 1B, 2B, and 8B datasets. 
With near-data processing and an accuracy-preserving hierarchy, \OurSys{} reduces communication rounds and payload sizes, making network latency negligible.
From 1B to 2B, the hierarchy maintains the same number of levels, so only the per-level data volume increases. The top-level index becomes larger, leading to slightly higher computation overhead due to more expensive graph traversal sequentially, which is executed sequentially and incurs random memory access.
In contrast, computation and data access in each lower level proceed in parallel across Index Store nodes.
Meanwhile, the number of accessed partition vectors stays roughly constant across these scales, causing disk I/O to remain flat.
At 8B, an additional hierarchy level is introduced. This structural change reduces the size of the top-level index, mitigating serial computation overhead. However, it also requires accessing one more level of data, which increases disk I/O due to the extra I/O round introduced by the deeper hierarchy.
In summary, the latency characteristics of \OurSys{} scale predictably with the hierarchy depth: computation is dominated by serialized top-level traversal, while disk I/O is driven by the number of levels.

\begin{figure*}[!t]
    \centering
    \subfigure[SPACEV centroids.]{
        \includegraphics[width=0.235\linewidth]{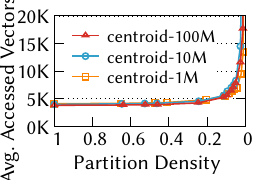}
        \label{fig:evaluation:sppspacev}
    }
    \subfigure[SIFT centroids.]{
        \includegraphics[width=0.235\linewidth]{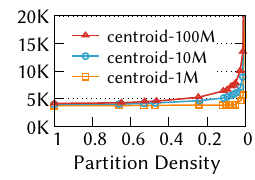}
        \label{fig:evaluation:sppsift}
    }
    \subfigure[DEEP.]{
        \includegraphics[width=0.235\linewidth]{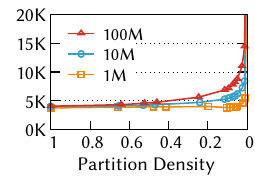}
        \label{fig:evaluation:sppdeep}
    }
    \subfigure[LAION.]{
        \includegraphics[width=0.235\linewidth]{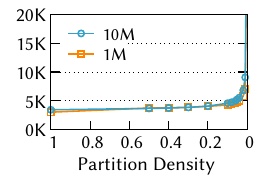}
        \label{fig:evaluation:spplaion}
    }
    \subfigure[TEXT.]{
        \includegraphics[width=0.235\linewidth]{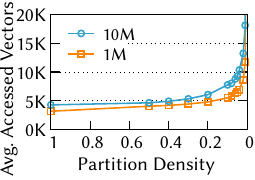}
        \label{fig:evaluation:spptext}
    }
    \subfigure[Cohere.]{
        \includegraphics[width=0.235\linewidth]{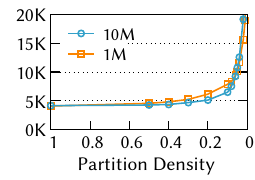}
        \label{fig:evaluation:sppcohere}
    }
    \subfigure[BIOASQ.]{
        \includegraphics[width=0.235\linewidth]{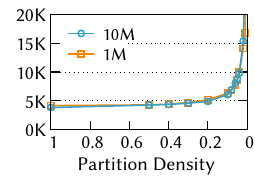}
        \label{fig:evaluation:sppBIOASQ}
    }
    \subfigure[OpenAI.]{
        \includegraphics[width=0.235\linewidth]{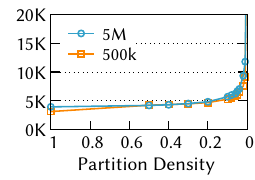}
        \label{fig:evaluation:sppopenai}
    }
    \caption{The number of accessed vectors to reach $recall@5=0.9$ under different partition densities across eight datasets spanning diverse scales. 
    The “centroid’’ refers to the cluster centroids produced under the balanced partition granularity.
All datasets exhibit a similar trend and consistently reveal the presence of a balanced granularity. }
    \label{fig:evaluation:spp}
\end{figure*}


\subsection{Simulation at Extreme Scale}
\label{sec:eval_simulator}
Due to the limitation of available datasets (up to 8B vectors), we simulate \OurSys{}’s performance at extreme scales, extending to 1024 billion vectors. The system’s behavior is amenable to modeling because the number of index levels is fully determined by the dataset size and the chosen partition granularity. Since the same granularity and search parameters yield stable performance from 1B to 8B vectors as demonstrated in \S\ref{sec:eval_goodscalability}, we assume they remain effective at larger scales. We then estimate \OurSys{}’s throughput based on the algorithmic complexity of index search and the constraints imposed by the underlying hardware. Our cost model incorporates CPU capabilities, network bandwidth, and latency, and both disk IOPS and bandwidth.

Figure~\ref{fig:evaluation:simulator} presents the simulated throughput of \OurSys{} on Lsv3 instances from 1B to 1024B vectors. Each node is provisioned with a 4GB memory budget, which bounds the size of the top-level index. \OurSys{} scales well with the dataset size -- and correspondingly, with the number of servers. For a 1024B vector dataset, a constrained memory budget of 4 GB necessitates a six-level hierarchy, yielding a simulated average latency of 16 ms. Increasing the memory budget to 512 GB allows the hierarchy to flatten to four levels, reducing average latency to 10 ms.

The simulated results on the 1B, 2B, and 8B data closely match the end-to-end throughput numbers in Figure~\ref{fig:evaluation:scalability}, with the difference variance up to $6\%$, proving the high accuracy of our cost model. 
The simulation reveals that disk IOPS is the dominant bottleneck governing end-to-end throughput; in contrast, network bandwidth and CPU cycles are underutilized, saturating at only ~30\% and ~50\%, respectively.
Because aggregating disk IOPS increases linearly with the number of servers, and the number of vectors probed per query grows logarithmically with dataset size, \OurSys{} maintains high throughput scalability even at trillion-scale deployments.

The above analysis assumes that requests are uniformly distributed across servers. In practice, workloads may exhibit skew. We model this by introducing a load-imbalance factor $\beta$, where the hottest server receives $\beta$ times the average traffic. Our end-to-end evaluation on the 8B dataset shows $\beta = 1.2$. Applying this factor to all simulated scales introduces only minor variation in throughput. Skewness can be mitigated using well-studied techniques such as caching~\cite{liu2019distcache} or replication~\cite{rashmi2016ec, li2020pegasus, suresh2015c3} at storage nodes, which are orthogonal to the design of \OurSys{}.

\begin{figure}[!t]
    \centering
    \includegraphics[width=\linewidth]{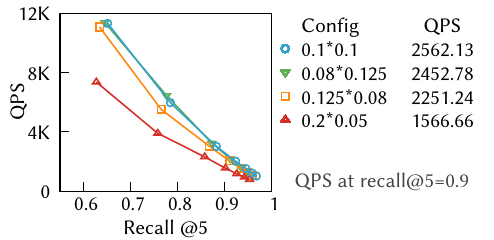}
    \caption{QPS--recall trade--off on SIFT100M using a three-level index with different density configurations ($level 0 \times level 1$). Alternative density configurations that deviate slightly from the balanced granularity (0.08 -- 0.125) exhibit performance comparable to the default setting, demonstrating that the system is relatively robust to granularity selection.
    }
    \label{fig:eval:granularity_benefits_tradeoff}
\end{figure}

\begin{figure*}[!t]
    \centering
    \includegraphics[width=1\linewidth]{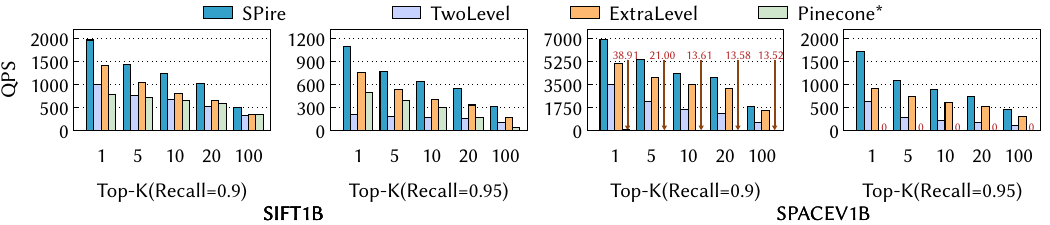}
    \caption{The effect of different hierarchy construction methods under different recall settings and datasets. \OurSys{} always delivers the best throughput.}
    \label{fig:eval:hierarchy_benefits}
\end{figure*}

\subsection{Existence of Balanced Granularity}
\label{sec:eval:balanced_granularity}
Now we examine whether balanced granularity consistently emerges across datasets and data scales.
This experiment covers the first eight datasets listed in Table~\ref{tab:datasets}, spanning text, image, and biomedical domains.
For each dataset, we measure the average search cost --- quantified as the number of accessed vectors required to achieve $Recall@5 = 0.9$ --- under varying partition densities.

Since Figure~\ref{fig:IndexSystemOverview} already presents results for SPACEV and SIFT, we use these two datasets to investigate whether balanced granularity also appears at upper index levels.
Figure~\ref{fig:evaluation:spp}(a–b) reports the results on centroids at different levels: ``centroid-100M'' corresponds to level-1 centroids, ``centroid-10M'' to level-2, and so on.
Across all levels, we observe a consistent turning point, indicating that balanced granularity persists throughout the hierarchy.
This finding supports the effectiveness of recursive indexing, where both accuracy and throughput remain stable across levels.

Figures~\ref{fig:evaluation:spp}(c–h) show results for the remaining datasets.
For each dataset, we vary the data scale by sampling different numbers of vectors from the base corpus.
Across all datasets and scales, balanced granularity consistently emerges, although the optimal value varies slightly across cases.
Overall, the results suggest that a partition density of 0.1 performs robustly in practice.

\subsection{Effect of \OurSys{} Design Choices}
\label{sec:eval_hierarchy_benefits}
This section quantifies the impact of \OurSys{}’s key design choices, focusing on balanced granularity and the accuracy-preserving hierarchy.

\noindent\textbf{Impact of partition granularity.} 
We first examine how different partition granularities affect the end-to-end throughput and recall of a three-level \OurSys{} on SIFT100M.
Our goal is to evaluate the sensitivity of \OurSys{} to granularity choices by comparing the balanced configuration (0.1) with three alternative density combinations ($level 0 \times level 1$).
The configurations $0.08 \times 0.125$ and $0.125 \times 0.08$ represent slight deviations from the default, whereas $0.2 \times 0.05$ constitutes a substantial departure.

As shown in Figure~\ref{fig:eval:granularity_benefits_tradeoff},
\OurSys{} achieves the highest throughput under the balanced $0.1\times0.1$ configuration, as this setting minimizes the total number of vector accesses.
Configurations with small deviations exhibit nearly identical performance, suggesting that the effective operating region for balanced granularity is relatively broad.
In contrast, the significantly divergent configuration ($0.2 \times 0.05$) incurs substantial throughput degradation.
These results highlight the importance of balanced granularity in achieving high performance and accuracy in \OurSys{}.

\noindent\textbf{Comparison with different hierarchy methods.}
We evaluate the effect of the accuracy-preserving hierarchy in \OurSys{} by comparing it with several alternative designs. We fix the memory budget at 4 GB and conduct experiments on SIFT1B and SPACE1B. To capture a broad range of application requirements, we vary $k$ in Recall@$k$ from 1 to 100. We compare the three-level hierarchy used in \OurSys{} for 1 billion vectors 
against four baselines: TwoLevel, ExtraLevel, and Pinecone*. 

TwoLevel enforces a two-level hierarchy by selecting a partition density of 0.01 to satisfy the memory budget.
ExtraLevel constructs a deeper, four-level hierarchy with density configuration of $0.5 \times 0.2 \times 0.1$ for $level 0\times level 1 \times level 2$.
Pinecone* builds its hierarchy in a top-down manner and chooses granularity based on the memory budget, without enforcing accuracy preservation.

Figure \ref{fig:eval:hierarchy_benefits} summarizes the results. On SIFT1B at Recall@10 = 0.9, \OurSys{} achieves 1.85$\times$, 1.52$\times$, and 1.89$\times$ higher throughput than TwoLevel, ExtraLevel and Pinecone*, respectively. This trend holds across different recall targets. TwoLevel underperforms because its coarse granularity increases the number of vectors that must be read. In contrast, ExtraLevel adds an unnecessary indexing level, which increases network round-trip time and disk I/O latency, ultimately bottlenecking end-to-end throughput. Pinecone* yields the lowest performance because its top-down construction does not preserve accuracy, requiring significantly more partition accesses to meet the accuracy constraint.

The advantages of \OurSys{} become even more pronounced on the skewed SPACE1B dataset, where 5–10\% of vectors are accessed by the majority of queries. At Recall@10 = 0.9, \OurSys{} delivers $2.77\times$ and 1.24$\times$ higher throughput than TwoLevel and ExtraLevel. Pinecone* performs especially poorly -- often producing near-zero throughput -- because its top-down hierarchy forces excessive partition reads. These results highlight the inefficiency of top-down iterative construction when handling skewed vector distributions.

\begin{figure}[!t]
    \centering
    \subfigure[SPACEV1B search cost with Accuracy.]{
        \includegraphics[scale=1]{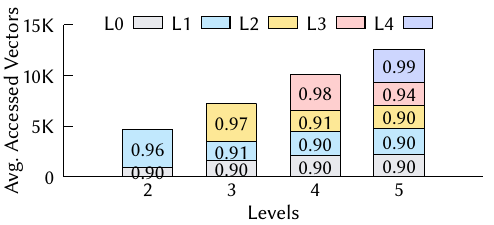}
        \label{fig:eval:searchcost:spacev}
    }
    \subfigure[SIFT1B search cost with Accuracy.]{
        \includegraphics[scale=1]{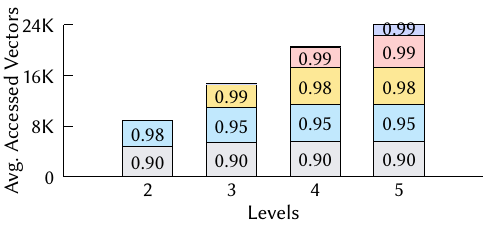}
        \label{fig:eval:searchcost:sift}
    }
    \caption{Search cost with accuracy of each level on 1B datasets. Adding a new level introduces a fixed search cost.}
    \label{fig:eval:searchcost}
\end{figure}

\subsection{Micro Analysis}
\label{sec:eval_ablationstudy}

\noindent
We next conduct a micro-analysis to characterize how the number of index levels influences search cost and resource consumption. 
Here, search cost is defined as the average number of vector entries accessed during search, and resource consumption refers to the storage and memory usage of the index at each level.
We also perform an ablation study to quantify the individual contributions of near-data processing and hash-based partition placement to the overall performance.


\subsubsection{Impact of Varying Number of Index Levels}
We set different memory budgets for the top-level index on the SPACEV1B and SIFT1B datasets, enabling \OurSys{} to construct indexes with varying numbers of levels. 

\noindent\textbf{Search cost at varying levels.} To guarantee a target accuracy of 0.90 at the bottom level (Level-0), upper levels must maintain significantly stricter accuracy thresholds (e.g., $>0.95$). Notably, this strict requirement does not incur prohibitive search costs but introduces fixed search costs. As illustrated in Figure~\ref{fig:eval:searchcost}, achieving 0.99 accuracy at the top level incurs a cost comparable to the search at the bottom level (e.g., 2,316 vectors for Level-4 to achieve 0.99 vs. 5,585 for Level-0 on SIFT1B to achieve 0.90). This trend is also validated on SPACEV1B.

\begin{figure}[!t]
    \centering
    \subfigure[Storage usage.]{
        \includegraphics[width=0.47\linewidth]{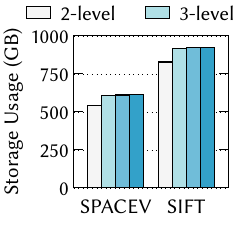}
        \label{fig:eval_levelUsage:disk}
    }
    \subfigure[Memory usage.]{
        \includegraphics[width=0.47\linewidth]{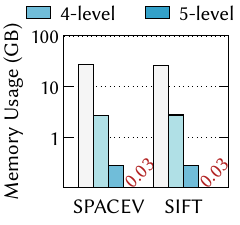}
        \label{fig:eval_levelUsage:memory}
    }
    \caption{Storage \& memory usage by index levels. Extra levels introduce negligible storage overhead while significantly reducing memory usage.}
    \label{fig:eval_levelUsage}
\end{figure}

\noindent\textbf{Resource consumption at varying levels.} Figure~\ref{fig:eval_levelUsage:disk} demonstrates that increasing the hierarchy depth imposes negligible storage overhead. With a partition density of 0.1, the bottom-level index occupies $C$ storage. Adding an additional level increases the total index size (excluding the top level) by only $C \cdot 10^{-(N-2)}$ for an $N$-level index. In contrast, the memory usage of the top-level index decreases significantly with more levels. As shown in Figure~\ref{fig:eval_levelUsage:memory}, with a partition density of 0.1, the top level stores only $10^{-(N-1)}$ of the total vectors, making it easier to fit into memory. These results demonstrate that multi-level indexing introduces minimal storage overhead while substantially reducing memory footprint. 

\begin{table}[!t]
\centering
\resizebox{\linewidth}{!}{%
\begin{tabular}{|l|l|l|l|l|l|}
\hline
MemoryBudget & 30GB & 3GB & 0.3GB & 0.03GB \\ \hline
Levels & 2   & 3 & 4 & 5 \\ \hline
Recall@10(SPACEV1B) & 0.927 & 0.913 & 0.905 & 0.902 \\ \hline
Latency(SPACEV1B) & 1.127 & 1.404 & 1.773 & 2.109 \\ \hline
Recall@10(SIFT1B)   & 0.909 & 0.902 & 0.901 & 0.901 \\ \hline
Latency(SIFT1B)   & 1.377 & 1.959 & 2.695 & 3.308 \\ \hline
\end{tabular}
}
\caption{Impact of increasing index levels on search accuracy and latency (ms) under different memory budgets (single-threaded, fixed parameters). }
\label{tab:eval_levelaccuracy}
\end{table}

\noindent\textbf{Latency at varying levels.}
We evaluate search latency among five Lsv3 VMs with a single-threaded client to quantify the latency of the same search parameter of increasing index depth. Our multi-level architecture demonstrates superior efficiency by strictly bounding communication costs. Table~\ref{tab:eval_levelaccuracy} shows that each additional level incurs a marginal latency penalty of approximately 0.3\,ms on \textsc{SpaceV1B} and 0.6\,ms on SIFT1B. This scaling allows for massive capacity within a constrained footprint: given a 32\,GB memory budget, a 4-level index supports 100 billion vectors, while a 5-level index scales to 1 trillion vectors. In both cases, the total latency remains well within the strict SLA bounds of online applications.

Notably, on the SPACEV1B dataset (Recall@5=0.9), our system outperforms the reported results of CXL-ANNS, despite the latter's use of high-speed CXL interconnects. This performance advantage stems from the inherent data skew of SPACEV1B. In traditional graph-based approaches (like CXL-ANNS), skew exacerbates the difficulty of nearest neighbor localization, forcing the traversal algorithm to perform excessive remote access hops. Our accuracy-preserving hierarchy circumvents this behavior, delivering lower latency without specialized hardware.

\begin{figure}[!t]
    \centering
    \subfigure[Different parameters.]{
        \includegraphics[width=0.47\linewidth]{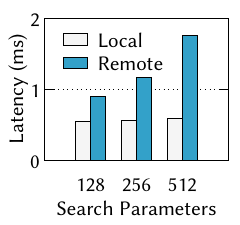}
        \label{fig:eval_neardata:partition}
    }
    \subfigure[Different levels.]{
        \includegraphics[width=0.47\linewidth]{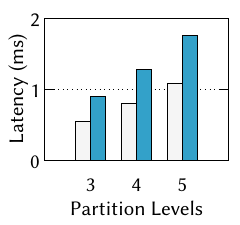}
        \label{fig:eval_neardata:level}
    }
    \caption{
    Query network latency for near-data processing (local) and remote raw vector transfer (remote). Near-data processing reduces the network overhead.
    }
    \label{fig:eval_neardata}
\end{figure}

\subsubsection{Ablation Study}

\noindent\textbf{Low network latency via near-data processing.} 
We quantify the benefits of near-data processing by comparing it against a baseline of raw vector transfer, varying both the search parameter ($N$, partitions read) and index depth.

For a three-level index, with \(N = 128\), \(256\), and \(512\), figure~\ref{fig:eval_neardata:partition} illustrates the effect of near-data processing. With the same search parameter, near-data processing reduces latency compared to raw vector transfer, primarily by decreasing network latency through smaller data transmissions. Specifically, latency is reduced by 1.62$\times$, 2.06$\times$, and 2.98$\times$ for \(N = 128\), \(256\), and \(512\), respectively. 
As \(N\) increases, the latency increases due to higher workload, but the impact on network latency remains minimal, as near-data processing significantly reduces the amount of transmitted data. Figure~\ref{fig:eval_neardata:level} shows latency improvements across different numbers of levels with a fixed \(N = 128\). Latency is reduced by 1.61$\times$ to 1.63$\times$ for varying levels.

\begin{figure}[!t]
    \centering
    \subfigure[Latency.]{
        \includegraphics[width=0.47\linewidth]{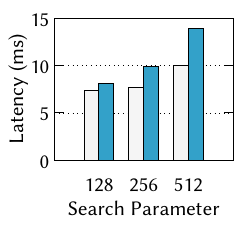}
        \label{fig:eval_localityInsensitive:partition:lat}
    }
    \subfigure[Throughput.]{
        \includegraphics[width=0.47\linewidth]{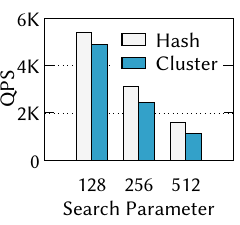}
        \label{fig:eval_localityInsensitive:partition:tps}
    }
    \caption{Impact of partition placement policies on latency and throughput.
    We compare the hash-based placement policy (used by \OurSys{}) with a cluster-based placement baseline. The hash-based policy distributes partitions uniformly across nodes, mitigating access hotspots and yielding higher throughput and lower latency.}
    \label{fig:eval_localityInsensitive}
\end{figure}

\noindent\textbf{Balanced access via hash-based partition placement.} We compare two partition placement strategies: (1) the hash-based placement used by \OurSys{}, which assigns partitions to nodes via a hash function over the partition IDs, and (2) a cluster-based baseline strategy that colocates partitions whose centroids are close in the vector space.
We evaluated the impact of these placement strategies on latency by varying the number of partitions \(N\) read at each level.  
Figure~\ref{fig:eval_localityInsensitive} shows peak throughput and average latency for different values of \(N\). As \(N\) increases, the cluster-based strategy creates hotspots, resulting in higher latency and lower throughput. 
In contrast, the hash-based placement distributes requests evenly, reducing latency by \(1.10\times\), \(1.28\times\), and \(1.40\times\) for \(N = 128\), \(256\), and \(512\), respectively, while increasing throughput by the same factors.


\section{Related Work}  
\noindent We critically review existing literature on Distributed ANNS, highlighting the fundamental, unresolved system trade-off established in ~\S\ref{background}. 

\noindent\textbf{Graph-based indexes}
Graph-based indexes, such as HNSW~\cite{Malkov2018TPAMI}, introduce core system conflicts in distributed settings. Their inherent dynamic and data-dependent traversal necessitates frequent, synchronous cross-node RPCs, imposing significant latency overhead.
Prior works have attempted to mitigate this bottleneck using hardware acceleration and storage-level optimizations. CXL-ANNS~\cite{Jang2023ATC} uses CXL-attached remote memory to distribute graph indexes, and CoTra~\cite{CoTra} leverages RDMA to access a sharded graph.
PipeANN~\cite{osdi25pipeann} accelerates single-node graph-based search by aligning best-first traversal with SSD access patterns.
While these approaches reduce latency in local or hardware-assisted environments, they do not eliminate the tight coupling of distributed graph traversal, where each hop still triggers remote access.
Hence, \OurSys{} focuses on addressing the fundamental system-design and mechanism-level challenges in distributed vector search, without relying on specialized hardware.

\noindent\textbf{Partition-based indexes}
In contrast, partitioning methods, such as Inverted File Index (IVF)~\cite{FAISS2020}, enable search parallelism by localizing query execution to a small subset of partitions, effectively decoupling the search dependency inherent in graph traversal. Distributed systems adopting this approach generally fall into two categories:
\textbf{Na\"ive partitioning:} Partitioning based on non-semantic criteria (e.g., random assignment or chronological order)~\cite{Suchal2010Full, huang2020kdd, zhang2018kdd, gao2024arxiv, Nara2014VLDB}. These methods often require traversing nearly all partitions for acceptable recall.
\textbf{Spatial partitioning :} Partitioning based on semantic similarity (e.g., k-means clustering) to target relevant subsets of partitions~\cite{Wei2020VLDB, ChenW21NIPS, Borgeaud2021ICML, Deng2019BigData, PineconeServerless, MILVUS2021SIGMOD}.
While this parallel design offers low latency, it still depends on high probing amplification to maintain recall, which severely restricts the maximum achievable Throughput (QPS).
HARMONY~\cite{xu2025harmony} combines dimension-based and vector-based partitioning to improve load balance across nodes, but cannot remove this amplification.
While optimized local systems such as ScaNN~\cite{scann} and Quake~\cite{mohoney2025quakeadaptiveindexingvector} successfully leverage CPU acceleration to optimize computational efficiency, they still incur substantial network and I/O overheads in distributed deployments due to the need to access large amounts of data.

\noindent\textbf{Optimizations for partition traversal.} 
Prior work reduces intra-partition cost through techniques such as LEAT~\cite{LEATSigmod20}, LEQAT~\cite{LEQATVLDBJ2023}, and Auncer~\cite{Zhang2023NSDI}, which use error bounds and early termination to prune traversal, while Vexless~\cite{su2024vexless} reduces the number of accessed partitions via tailored sharding.
These methods improve efficiency but still leave substantial work on irrelevant vectors within selected partitions. They are orthogonal to \OurSys{} and can be incorporated to further improve performance.


\section{Conclusion}

This paper presented \OurSys{}, a scalable hierarchical vector index that resolves tension between accuracy, latency, and throughput in distributed ANNS. \OurSys{} introduces two key techniques: a balanced partition granularity that prevents read-cost explosion, and an accuracy-preserving recursive construction that builds a multi-level index with predictable per-level search cost and stable end-to-end accuracy. As a result, \OurSys{} achieves high throughput scalability and substantial performance gains over state-of-the-art systems, reaching up to 9.64× higher throughput on production-scale workloads with billions of vectors. Its stateless design further simplifies deployment and enables elastic scaling. We hope \OurSys{} provides a foundation for the next generation of large-scale vector search systems.

\bibliographystyle{plain}
\bibliography{ref}

\clearpage

\end{document}